\documentclass[draftcls,onecolumn]{IEEEtran}

\usepackage{epsfig}

\begin{document}
\title{On the possibility of tunable-gap bilayer graphene FET}

\author{\normalsize Gianluca Fiori, Giuseppe Iannaccone\\
 Dipartimento di Ingegneria dell'Informazione~: Elettronica, Informatica, Telecomunicazioni,\\
Universit\`a di Pisa, Via Caruso, 56126 Pisa, Italy.\\
email : g.fiori@iet.unipi.it; Tel. +39 050 2217596; Fax : + 39 050 2217522\\}
\maketitle
\newpage
\vspace{5cm}
\begin{abstract}

We explore the device potential of tunable-gap bilayer graphene FET
exploiting the possibility of opening a bandgap in bilayer graphene
by applying a vertical electric field via independent gate operation.
We evaluate device behavior using atomistic simulations based on
the self-consistent solution of the Poisson and Schr\"odinger equations
within the NEGF formalism. We show that the concept works, but
bandgap opening is not strong enough to suppress band-to-band
tunneling in order to obtain a sufficiently large $I_{\rm on}/I_{\rm off}$ ratio
for CMOS device operation. 

\end{abstract}

{\bf Keyworks:} NEGF, graphene, bilayer, tight-binding Hamiltonian, Poisson/Schr\"odinger.
\newpage

\section{Introduction}

Graphene has entered the nanoelectronics scenario only recently and 
it is intensely investigated as one of the most promising candidates to replace
silicon as a channel material in nanoscale transistors.
Even though graphene is a gapless material, a significant energy gap can be opened in different ways,
like for example by ``rolling'' it in carbon nanotubes~\cite{Ijima} or by defining narrow graphene stripes
through electron beam lithography~\cite{Avouris} or chemical reaction~\cite{Dai}.
However, several unsolved technological problems arise: state of the art technology cannot indeed 
 conveniently control the chirality and the nanotube position, as well as define
 graphene nanoribbons with widths close to 1-2 nm with atomically flat edges,
in order to obtain a reasonable semiconducting energy gap.

Recently, theoretical models~\cite{Nilsson} and experiments~\cite{Ohta,Oostinga}
have shown that bilayer graphene has an energy gap controllable by a vertical electric field.
One could exploit this property to use bilayer graphene as a channel material for FETs, 
 defining an energy gap only when really needed, i.e. when the FET must be in
the off state. This capability could open the possibility of patterning a bilayer graphene sheet 
with lithographic techniques and non prohibitive feature sizes.

In this letter, we want to assess the possibility of fabricating a tunable-gap Bilayer Graphene (BG)
FET with independent gate operation, based on atomistic numerical simulations.
 We assume that all the technological challenges
associated to the reliable fabrication of bilayer graphene FETs can be solved, and evaluate
the potential performance of near ideal device structures. In our opinion, this is one of the
most powerful uses of computer simulations in guiding and orienting nanoelectronics research.

A semi-analytical model based on the effective mass approximation
has been presented in order to compare bilayer against monolayer graphene
transistors~\cite{Guo}. However, in that work a constant induced energy bandgap is assumed for 
bilayer graphene, missing the most important material property of the material.

For a physically sound analysis of BG-FETs, we have developed a code 
based on the Non-Equilibrium Green's Function formalism (NEGF), with a tight-binding Hamiltonian
on a $p_z$ orbital basis set in the real space, which has been included in our in-house device 
simulator NANOTCAD ViDES~\cite{ViDES}. 

We will show that BG-FETs miss the ITRS requirements~\cite{ITRS} for the $I_{\rm on}/I_{\rm off}$ ratio ($> 10^{4}$)
by a large amount, since the induced
gap is not sufficient to suppress band-to-band tunneling currents.

\section{Physical model and Results}

Our approach is based on the self-consistent solution of the 
Poisson and Schr\"odinger equations within the NEGF formalism
and the ballistic transport assumption, as described in~\cite{noi}.

The considered bilayer graphene Hamiltonian is composed by the two
single layer graphene Hamiltonians, coupled by the $t_p$=0.35~eV hopping parameters 
in correspondence of overlaying atoms along the 
$z$-direction~\cite{Nilsson}: the elementary cell is depicted in Fig.~\ref{Fig1}a.
 Semi-infinite contacts have been modeled
along the $y$ direction by means of self-energies, while periodic boundary conditions 
have been imposed in the $x$ direction, with period equal to 
$\sqrt{3}a_{cc}$, where $a_{cc}=0.144$~nm is the carbon-carbon bonding distance.

The simulated device is a double-gate BG-FET, whose structure is shown in Fig.~\ref{Fig1}b.
We assume metal gates, and a 1.5~nm layer of SiO$_2$ as gate dielectric. We also assume 
an air gap of 0.5~nm between the dielectric interface and the position of carbon sites, 
as evaluated in~\cite{Dai}. The channel is 15~nm long, and
the inter-layer distance is 0.35~nm. The source and drain extensions are 10~nm long, 
and are doped with an equivalent molar fraction of fully ionized donors $f_d=5\times10^{-3}$. 

Particular attention has to be posed in the computation of the mobile charge. 
Considering for the sake of simplicity the equilibrium case, the mobile charge on
a given site on layer $j$ ($j=1,2$) of the bilayer graphene sheet reads
\begin{eqnarray}
\rho_j &=&-2q \int_{E_i}^{+\infty} dE \; LDOS_{j} (E) f(E) \nonumber \\
&& +2q\int_{E_i}^{+\infty} dE \; LDOS_{j}(E)\left[ 1-f(E)\right]
\label{rho}
\end{eqnarray}
where $q$ is the elementary charge, $E_i$ is the intrinsic (mid-gap) Fermi level, 
$f$ is the Fermi-Dirac occupation factor, and $LDOS_{j}(E)$ is the local density of states
on layer $j$.

Let us stress the fact that, in order to avoid an unphysical antiscreening behavior, 
it is necessary to include the effect of dielectric 
polarization along the direction perpendicular to the graphene bilayer, due to the response
of valence electrons to the applied field.
We include this effect by introducing a polarization factor $\alpha$ defined as 
\begin{equation}
\alpha(V)=\frac{\int_{-\infty}^{E_i}LDOS_{1}(E,V)dE}{\int_{-\infty}^{E_i}\left[ LDOS_{1}(E,V)+LDOS_{2}(E,V)\right] dE},
\label{alpha}
\end{equation}
where $V$ is the potential difference between the two graphene layers.
By using $\alpha$, we can write the total charge (mobile charge, valence band electrons and ions) 
on a given site of layers 1 and 2 as 
\begin{eqnarray}
\rho^\prime_1&=&\rho_1+q[1-2\alpha(V)] \nonumber \\
\rho^\prime_2&=&\rho_2-q[1-2\alpha(V)] .
\end{eqnarray}

However, performing the integral from $-\infty$ to $E_i$ in (\ref{alpha}) 
at each iteration step can be too computationally demanding.
To this purpose, we compute the polarization factor as a function of $V$ only once, 
for an infinitely long BG-FET (solid line in Fig.~\ref{Fig1}c).
As can be seen, $\alpha(V)$ can be reasonably approximated by a straight line 
$\alpha(V) = 0.5 + 3.282 \times 10^{-3} V$. In our code, we have used an odd polynomial of
11th order obtained through least mean square fitting (symbols in Fig.~\ref{Fig1}c).

In Fig.~\ref{Fig2} the transfer characteristics for $V_{DS}=0.5~V$ are shown, for different gate configurations.
In particular, in Fig.~\ref{Fig2}a the transfer characteristics as a function of the top gate voltage $V_{\rm top}$ 
for two bottom gate voltages ($V_{\rm bottom}=0~V$ and $V_{\rm bottom}=-1.0~V$) are shown.
As can be seen, large currents can be obtained above threshold, while the BG-FET shows huge problems 
to switch off. Even after applying really negative gate voltages, the drain-to-source current 
decreases only by roughly six times. The ratio even worsens when $V_{\rm bottom}=-1.0~V$,
since in this case a larger amount of positive charge is injected
in the valence band, due to band-to-band tunneling, which pins the potential along the channel, 
and further degrades the gate control over the barrier.

Such results are close to those obtained in experiments on monolayer graphene devices~\cite{Lemme},
showing that the tunable gap is too small to effectively suppress band-to-band tunneling in the off state.

In order to increase the vertical electric field between the two layers of graphene,
we have driven the device by imposing a fixed voltage between the two gates 
$V_{\rm diff}=V_{\rm top}-V_{\rm bottom}$.

Even in this case, device transconductance (i.e. the slope of the transfer characteristic)
almost remains the same, regardless of the applied $V_{\rm diff}$.
This can be explained by the large accumulation of positive charge in the channel, caused by 
band-to-band tunneling in correspondence of the drain~\cite{noi}: this charge
screens the vertical electric field, limiting the size of the opened bandgap.

In Fig~\ref{Fig3}b, the transmission coefficient as a function of the transversal 
wave vector $k_x$ is shown for $V_{top}=-4.0~V$, $V_{DS}=0.5~V$ and $V_{bottom}=0~V$, i.e., close to minimum
achievable current, and in correspondence of the band minimum. Different energy intervals are identified with letters
and compared to a sketch of the band edge profile in Fig.~\ref{Fig3}a 
as an aid to interpretation. As can be seen, energy intervals 
tagged as $B$, $D$ and $F$ correspond to the bandgaps in the channel, source and drain extensions, respectively:
the gap is not uniform, it is maximum in the channel, where it is almost 100 meV. One can also see the
bound states in the valence band in the energy interval $C$ and the valence band states in $E$, which
are populated by holes for negative voltages, and which eventually pin the potential of the two mono-layers of graphene, i.e. 
reducing the gate effectiveness in controlling the channel barrier and the bandgap.

Band-to-band tunneling can in principle be reduced by engineering the electric field between source and channel.
Here we make two attempts, by introducing a lateral spacer on the gate sides of 2.5~nm,  
and by introducing a linear doping profile degrading to zero in 2.5~nm under the spacer. Both options 
have a positive but limited effect, as shown in Fig. 3c, and do not solve the problem.

\section{Conclusions}

We have explored the possibility of realizing FETs by exploiting the tunable-gap
property of bilayer graphene, through a code based on 
Tight-Binding NEGF device simulations. Aiming at technology foresight,
we have considered the overly optimistic 
case of an ideal structure and of ballistic transport.

Our computer simulations show that tuning the gap is not a very promising technique
for achieving appropriate switching characteristics of BG-FETs, due to the 
too weak suppression of band-to-band tunneling. A more complete exploration of the 
design space, in terms of structure and bias, would be required to cast a 
definitive answer, that is beyond the scope of this letter.

\section*{Acknowledgment}

The work was supported in part by the EC Seventh Framework
Program under project GRAND (Contract 215752), by the Network of Excellence
NANOSIL (Contract 216171), and by the European Science Foundation
EUROCORES Program Fundamentals of Nanoelectronics, through funds from
CNR and the EC Sixth Framework Program, under project DEWINT (ContractERAS-CT-2003-980409).

\newpage

\newpage


\begin{figure}
\caption{a) Elementary cell of the simulated bilayer graphene; b) transversal cross-section of the simulated 
graphene bilayer field effect transistor; c) tight-binding (solid line) and least mean square fitting analytical 
results (symbols) for the polarization factor $\alpha$ in eq.~(\ref{alpha}). }
\label{Fig1}	
\caption{a) Transfer characteristics computed as a function of the top gate voltage, when the bottom gate
voltage $V_{bottom}$=0~V (solid line), and $V_{bottom}$=-1.0~V (dashed line); b) transfer characteristics computed as
a function of $V_{bottom}$ for $V_{diff}$=1.0~V (solid line) 
and $V_{diff}$=0.5~V (dashed line).}
\label{Fig2}	
\caption{a) Sketch of the band edge profile; 
b) Transmission coefficient as a function of the energy and the transversal wave vector $k_x$ for 
$V_{bottom}$= 0~V,  $V_{top}=-4.0~V$ and $V_{DS}=0.5~V$; b) transfer characteristics
for a BG-FET with 2.5~nm spacer, and a BG-FET with doping profile linearly varying over a 25~nm 
long region from 0 to $f_d$.}
\label{Fig3}	
\end{figure}


\begin{figure}
\begin{center}
\vspace{8cm}

\epsfig{file=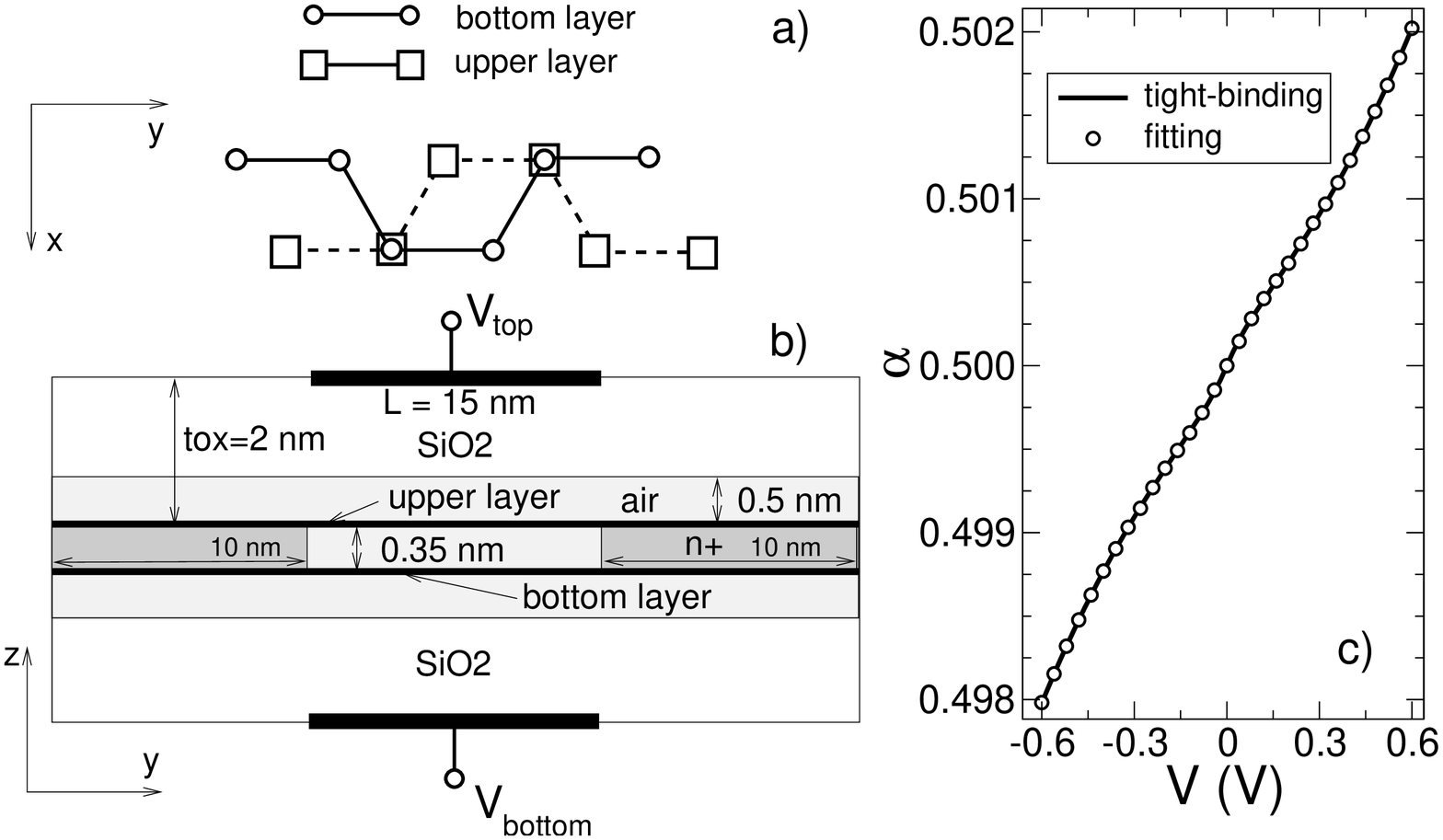,width=12cm} 
\huge
\vspace{1cm}\\
FIG. 1\\
Gianluca Fiori, Giuseppe Iannaccone \\
\normalsize	
\end{center}
\end{figure}

\begin{figure}
\begin{center}
\epsfig{file=./Fig2.eps,width=14cm} 
\huge
\vspace{1cm}\\
FIG. 2\\
Gianluca Fiori, Giuseppe Iannaccone \\
\normalsize	
\end{center}
\end{figure}

\begin{figure}
\begin{center}
\epsfig{file=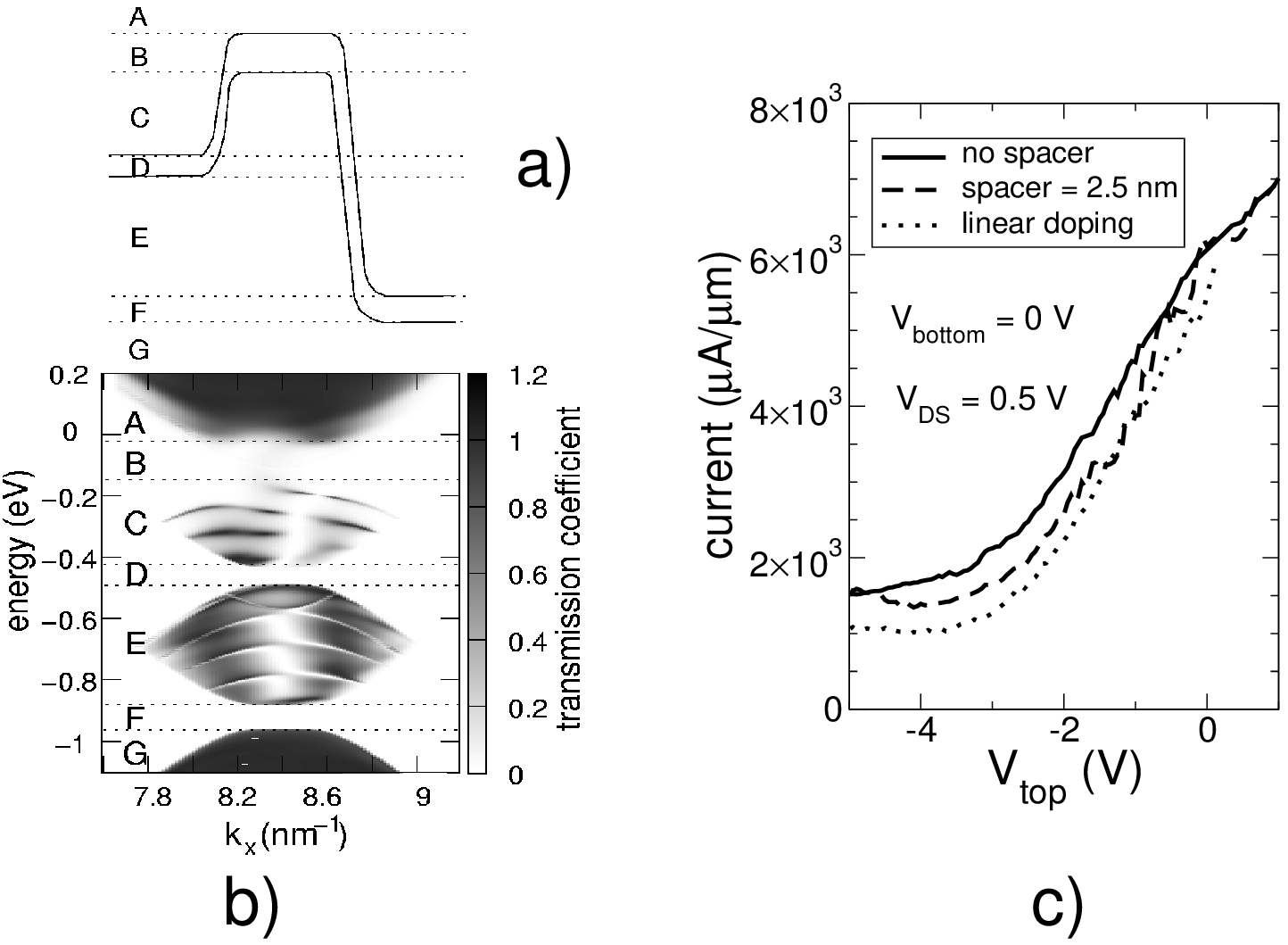,width=12cm} 
\huge
\vspace{1cm}\\
FIG. 3\\
Gianluca Fiori, Giuseppe Iannaccone \\
\normalsize	
\end{center}
\end{figure}

\end{document}